# Evaluation of a meta-analysis of air quality and heart attacks, a case study

S. Stanley Young[a] and Warren B. Kindzierski[b,c]

[a]CGStat, Raleigh, NC, USA; [b]School of Public Health, University of Alberta, Edmonton, Alberta, Canada

[c]Corresponding author: School of Public Health, University of Alberta, 3-57 South Academic Building, 11405-87 Avenue, Edmonton, Alberta, T6G 1C9 Canada; warrenk@ualberta.ca.

# Evaluation of a meta-analysis of air quality and heart attacks, a case study

ABSTRACT

It is generally acknowledged that claims from observational studies often fail to replicate. An exploratory study was undertaken to assess the reliability of base studies used in meta-analysis of short-term air quality-myocardial infarction risk and to judge the reliability of statistical evidence from meta-analysis that uses data from observational studies. A highly cited meta-analysis paper examining whether short-term air quality exposure triggers myocardial infarction was evaluated as a case study. The paper considered six air quality components - carbon monoxide, nitrogen dioxide, sulphur dioxide, particulate matter 10 μm and 2.5 μm in diameter (PM10 and PM2.5), and ozone. The number of possible questions and statistical models at issue in each of 34 base papers used were estimated and p-value plots for each of the air components were constructed to evaluate the effect heterogeneity of p-values used from the base papers. Analysis search spaces (number of statistical tests possible) in the base papers were large, median = 12,288 (interquartile range = 2,496−58,368), in comparison to actual statistical test results presented. Statistical test results taken from the base papers may not provide unbiased measures of effect for meta-analysis. Shapes of p-value plots for the six air components were consistent with the possibility of analysis manipulation to obtain small p-values in several base papers. Results suggest the appearance of heterogeneous, researcher-generated p-values used in the meta-analysis rather than unbiased evidence of real effects for air quality. We conclude that this meta-analysis does not provide reliable evidence for an association of air quality components with myocardial risk.

**Introduction**

**Meta-analysis**. Often it is not possible to conduct a randomized clinical trial, so meta-analysis (MA) of existing observational studies is taken as a valid way in many scientific fields, including the biomedical sciences, to summarize and address a common research question. MA gathers multiple papers that address a common research question and takes a statistical estimate from each paper to feed into an analysis process. MA is intended to overcome a number of problems associated with traditional literature or systematic reviews (Wolf 1986):

- Selective inclusion of studies, often based on a reviewer's own impressionistic view of the quality of the study.
- Differential subjective weighting of studies in the interpretation of a set of findings.
- Misleading interpretations of study findings.
- Failure to examine characteristics of the studies as potential explanations for dissimilar or consistent results across studies.
- Failure to examine moderating variables in the relationship under examination.

A key assumption of MA is that an estimate coming from an observational study is unbiased and fair estimate of the research question (Boos and Stefanski 2013). Another important assumption is that MA of multiple studies offers a pooled estimate with increased precision (Cleophas and Zwinderman 2015). MA is also known for not being able to resolve all potential problems associated with building reliable and valid knowledge from scientific literature focussing on a common research question. Glass et al. (1981) summarized limitations of MA into four categories:

- Logical conclusions cannot be drawn by comparing and aggregating studies in MA that include different measuring techniques, definitions of variables (e.g., treatments, outcomes), and subjects because they are too dissimilar.
- Results of MA are uninterpretable because results from "poorly" designed studies are included along with results from "good" studies.
- Published studies are biased in favour of significant findings because nonsignificant findings are rarely published and this in turn leads to biased MA results.
- Multiple results from the same study are often used which may bias or invalidate the MA and make the results appear more reliable than they really are, because these results are not independent.

**P-hacking**. It is becoming known that the reliability of observational studies can be poor (Hubbard 2015, Chamber 2017, Harris 2017). One cause of this is asking a lot of questions and testing numerous statistical models, or multiple testing and multiple modelling (MTMM), without any statistical correction – so called p-hacking (Ellenberg 2014, Hubbard 2015, Chamber 2017, Harris 2017). At best such work is exploratory and at worst it represents inconsistent (flawed) statistical analysis (Glaeser 2006). Inconsistent statistical estimates from studies that are exploratory are not reliable for inclusion in MA as they cannot be taken to be unbiased. Head et al. (2015) examined evidence of p-hacking in published studies by doing a text-mining search of all Open Access papers in the PubMed database and concluded:

> "Our text-mining suggests that p-hacking is widespread… across all scientific disciplines for which data are available…" and "…supports the conclusion that p-hacking is rife."

Bayesian Model Averaging (BMA – application of Bayesian inference to model selection, combined estimation and prediction of variable effects) may be immune from the criticism of p-hacking. It has been promoted as a way to estimate air quality−health effects that account for model uncertainty (e.g., Clyde 2000, Koop and Tole 2004, Koop et al. 2010). However, BMA can over- or under-estimate health effects, depending on correlations of the variables involved (Dominici et al. 2008), and there is a known relationship between Bayes factors and p-values (Held and Ott 2018). Another averaging approach, bootstrap model averaging (Wang et al. 2015a), incorporates model uncertainty that results from searching through a set of candidate models and results are obtained easier than through Bayesian analysis (Hastie et al. 2009). In reality, any type of averaging requires an assumption that results are at least loosely homogeneous. If one is dealing with a mixture – i.e., heterogeneous results, averaging makes no sense.

Interestingly, both BMA and bootstrap model averaging in air quality−health effect studies suggest the possibility that more particulate matter in air leads to fewer acute deaths (Koop et al. 2010) and MI hospitalizations (Wang et al. 2015a). While not solely an issue with BMA and bootstrap model averaging, this is implausible from a dose−response point-of-view. This speaks more to the overall challenges and limitations of current observational epidemiology methods for studying weak risk factors for chronic diseases in our population when important risk factors for these diseases are unaccounted for.

**Study objectives**. Valid concerns exist today in the body of published observational studies for which biomedical researchers and policy makers should be aware of:

(1) MTMM can give rise to false positive results (Westfall and Young 1993).

(2) Analysis manipulation, intentional or unintentional, can give rise to small p-values so the contention that replication implies a correct claim is questioned (Simonsohn et al. 2014).

(3) As negative effect studies are generally more difficult to publish than positive effect studies (Hubbard 2015, Chamber 2017, Harris 2017), over time there are many more positive studies in scientific literature than negative studies, so a false positive effect claim can mistakenly become established fact (Nissen et al. 2016).

Ambient air quality is of interest to public health and today a large body of air quality−health effect observational studies exists in published literature. Many of these studies support the current paradigm that air components of public health concern are causal of effects such as death and myocardial infarction (MI) from short-term exposure. Specific to these endpoints there also exists, to a lesser extent, published studies that show no effect from short-term air quality exposure. Citations for a number of these studies are presented online in a background information report called "Background information for meta-analysis evaluation" in the Cornell University Library e-print service repository arXiv.org (Young and Kindzierski 2018, Info 02). Access to this background information is available for free to anyone at https://arxiv.org/abs/1808.04408. A number of these no-effect (negative) studies were based on large population samples and this evidence cannot be discounted. For this reason and concerns identified above we were interested in examining a MA of air components showing an effect on MI. We wanted to undertake an exploratory analysis of MA using accepted statistical methods nontraditional to environmental epidemiology in order to:

- Assess the reliability of base studies used in MA examining whether short-term exposure can trigger MI.
- Judge the reliability of statistical evidence from MA that uses air quality−health effect observational base studies.

It is well-known that claims coming from observational studies most often do not replicate. We contend that a large part of the failure to replicate is due to MTMM in the base papers with no statistical adjustment. We show that the results from the base papers can be viewed as a two-component mixture, some base papers indicate positive effects, where as other appear without effect. The fact that recent large studies indicate no air quality–MI effect (e.g., Tsai et al. 2012, Milojevic et al. 2014, Talbott et al. 2014, Wichmann et al. 2014, Wang et al. 2015b, Young et al. 2017) support that these negative studies are correct and positive studies are consistent with so-called p-hacking.

**Materials and methods**

Young et al. (2018) recently analyzed the reliability of 14 observational epidemiology base studies about particulate matter−MI effects that were combined in a MA published in The Lancet (Nawrot et al. 2011). In the current study, we wanted to identify and analyze a well (highly)-cited MA of air quality−MI effects. Using procedures described in our Supplement, we selected a single MA (Mustafic et al. 2012) that is considered a "Highly Cited Paper" by the online subscription-based scientific citation indexing service *Web of Science* for further investigation as our case study. As of April 2018 the *Web of Science* indicated that this study had received enough citations to place it in the top 1% of the academic field of Clinical Medicine for the publication year 2012. The effect of short-term exposure of six air components on MI was analyzed in the case study – carbon monoxide (CO), nitrogen dioxide ($NO_2$), sulphur dioxide ($SO_2$), and particulate matter 10 μm and 2.5 μm in diameter, PM10 and PM2.5 respectively, and ozone and a conclusion was (Mustafic et al. 2012):

> "All the main air pollutants, with the exception of ozone, were significantly associated with a near-term increase in MI (heart attack) risk."

A total of 34 base papers were used in our case meta-analysis study (refer to our Supplement where citations for these base papers are listed and a descriptive summary of each paper is provided in Table S1). Effect estimates (risk ratios, RRs) along with confidence limits for the six air quality components in the case study are provided in Table 1. RRs listed in Table 1 are all close to 1.000. If confidence limits do not include 1.000, claims of an effect can be taken as true; however, it is acknowledged that any sort of bias can give rise to small deviations from 1.000 (Young 2008, Federal Judicial Center 2011).

*Data*

Electronic copies of the 34 base papers were obtained and after reviewing, outcomes, predictors, covariates and lags in each paper were counted. Researchers studying air quality and effects look at air quality (e.g., air components discussed above) and environmental conditions (e.g., conditions such as temperature, wind speed, relative humidity or dew point temperature, etc.) on the event (MI) day and previous days (lags). The inference is that these components might induce MI some days after a short-term elevated exposure day.

Different averaging times may to be used to represent an air component predictor (e.g., 24-hour average, 6-hour average for hours 07:00 to 10:00 and 17:00 to 20:00, 12-hour average for hours 07:00 to 19:00, daily 1-hour maximum, daily 1-hour minimum, etc. value (Wang and Kindzierski 2015a)) or a predictor of environmental conditions (e.g., daily average, daily minimum, daily maximum, etc. value for air temperature (Wang and Kindzierski 2015b)). Researchers can examine many or all of these predictors and then only report those predictors that give the strongest positive effect. This potentially leads to large numbers of statistical tests that are unreported in a study (specifically, test results that show weak or negative effects).

Generally, environmental epidemiology studies seek large sample sizes, leading to very small estimates of experimental error. Even here large sample sizes do not protect against bias. Further, environmental epidemiology researchers essentially never adjust their statistical analysis for MTMM. None of the 34 base papers used in our case study adjusted their analysis for MTMM.

*Methods*

Our methods are non-traditional to mainstream environmental epidemiology and can be used to assess the reliability of base studies used in MA. Initially, the 34 base studies were evaluated via simple counting; the resulting counts were used to approximate the analysis search space; which we define as the number of statistical tests possible in each study based on the possible outcomes, predictor variables, covariates and lag days that may have been used in statistical models to test for an effect. For example, cardiovascular outcomes might be presented in various studies as total cardiovascular disease, cardiac failure, ischemic heart disease, myocardial infarction, ST segment elevation myocardial infarction (STEMI), non-ST segment elevation myocardial infarction (NSTEMI), etc. Base papers with large analysis search spaces suggest the use of a large number of statistical models and statistical tests for an effect thereby allowing greater flexibility of researchers to selectively search thru and only report results showing positive effects. We counted the number of outcomes, predictors, covariates, etc. available in each base paper (covariates can be elusive as they might be mentioned anywhere in a paper). The search spaces were computed as follows:

- The product of outcomes, predictors and time lags = number of questions at issue, Space1.
- A covariate can be in or out of a model, so one way to approximate the modelling options is to raise 2 to the power of the number of covariates, Space2.
- The product of Space1 and Space2 = an approximation of analysis search space, Space3.

For this study we acknowledge that this approximation of analysis search space essentially represents a lower bound. As to why this is, we previously discussed that there exists many, many possible different averaging times that can be used to represent an air component or a temperature predictor, potentially leading to large analysis search space numbers. In addition, researchers have used logarithmic transformations to further represent air component concentrations in air quality−effect studies. For example, Krewski et al. (2009) employed the natural logarithm transformation of PM2.5 measurements in an air quality−mortality effects study. Ginevan and Watkins (2010) report that such transformations can be problematic in a number of ways: (1) where low air quality dose−mortality risks may be substantially overstated if the observed log (dose) fit is due to extraneous factors like dosimetric error or confounding, and (2) when making causal inferences. With regard to the second point, it is simply not established in scientific literature that analyses based on a logarithmic transformation should be taken as evidence of dose when making causal inferences of air quality effects.

A p-value plot after Schweder and Spjøtvoll (1982) was used to inspect the distribution of the set of p-values reported in our case study. The p-value can be defined as the probability, if nothing is going on, of obtaining a result equal to or more extreme than what was observed. The p-value is a random variable derived from a distribution of the test statistic used to analyze data and to test a null hypothesis. Under the null hypothesis, the p-value is distributed uniformly over the interval 0 to 1 regardless of sample size (Hung et al. 1997). "Nothing-is-going-on" is a statistical straw-man argument. If a p-value is sufficiently small, then the straw-man is defeated, and it is concluded that the observed result is not due to chance. A distribution of true null hypothesis points in a p-value plot should form a straight line (Schweder and Spjøtvoll 1982). The plot can be used to assess the validity of a false claim being taken as true and, specific to our interest, served to examine the reliability of base studies used in MA.

A p-value plot can be constructed and interpreted as follows (Schweder and Spjøtvoll 1982):

- p-values are ordered from smallest to largest and plotted against the integers, 1, 2, 3, …
- If the points on the plot follow an approximate 45-degree line, then the p-values are assumed to be from a random (chance) process.
- If the shape of the points is roughly a hockey stick (blade on the bottom left hand corner, shaft towards the top right hand corner), then those p-values on the blade are unlikely due to chance.

For example, if many foods are evaluated in a nutritional study for an association with an effect and p-values from tests of those foods follow a 45-degree line in a p-value plot, then chance rules (Young et al. 2009), whereas p-values on the blade of the hockey stick in a p-value plot may be real or due to p-hacking. In this case it may also be useful to examine its statistical reliability. A valuable statistical tool for this purpose is a volcano plot (Cui and Churchill 2003). Here, the negative of "the log base 10 of the p-value corresponding to a statistical test is plotted against the calculated effect size. P-values that spew high left and right have small p-values and large effects. A volcano plot facilitates seeing important effects in the context of all the comparisons at issue. A volcano plot can present a complicated picture when experimental error, magnitude of the reported effect and possible analysis manipulation varies across the studies.

The utility of using p-value and volcano plots for interpreting statistical effects is illustrated in our background information report (Young and Kindzierski 2018, Info 04) for a recently-published negative air quality−MI effect paper (Milojevic et al. 2014). Risk ratio and confidence limit data displayed in two of their illustrations (Figures 1 and 2) was requested from the authors, but not secured. Here we used the program WebPlotDigitizer, https://automeris.io/WebPlotDigitizer/, to capture the data displayed in these illustrations. P-values were computed for all RRs and confidence limits using a SAS JMP Add-In program (available on request). Our interpretation of resulting p-value and volcano plots constructed for the Milojevic et al. (2014) study is consistent with their finding that the air quality−MI effect is negative (refer to Young and Kindzierski 2018, Info 04).

**Results**

*Counting*

Counts of outcomes, predictors, covariates and lags for the 34 papers are given in Table 2. In each of the papers there are many thousands of possible analysis options. Summary statistics of the numbers of options are given in Table 3. Across the papers, the median number of possible analyses is 12,288 (interquartile range 2,496−58,368) for Space3, which takes all the factors into account.

*P-values*

The p-values for each of the air components in each base paper are given in Table 4. A blank cell indicates that that a base paper did not report a p-value on that air component. Only one base paper (citation 20, refer to our Supplement), examined all air components. Another base paper (citation 29), examined components of PM2.5, finding no significant effects when multiple testing was considered. The smallest/largest p-value for each of the air components are: ozone: 0.001/0.78; CO: 0.001/0.95; NO2: 0.001/0.79; SO2: 0.001/0.99; PM10: 0.001/0.90; PM2.5: 0.001/0.66. There are many small p-values reported. Looking down each column in Table 4, in addition to the small p-values, there are multiple p-values larger than 0.05. Researchers often take a p-value of 0.001 as virtual certainty. If a result has a p-value small enough to indicate certainty, then there should be few p-values larger than 0.05 (Boos and Stefanski 2011, Johnson 2013).

*P-value plots*

Figure 1 gives p-value plots for each of the six air components. For each air component, significant heterogeneity is apparent. Taking ozone as an example (upper left image), a p-value is often considered nominally statistically significant if it is equal to or smaller than 0.05. Nineteen p-value results are reported for ozone in Table 4 Many of the p-values are small, less than 0.05. But many are greater than 0.05. The pattern for ozone is like a hockey stick. The blade includes small p-values, which could be real effects or due to p-hacking. The p-values on the handle are consistent with random findings (if nothing is happening the plot of ranked p-values versus integers indicates a random, uniform distribution going left to right at 45-degrees). There is a gap in the upper right of the image plot for ozone, which might be due to chance or a file drawer effect.

What should we see if a risk factor (in this case an air component) is truly associated with (causes) heart attacks? There should be a preponderance of p-values below the 0.05 line. If there is a uniform effect of the air component on heart attacks, we should see a linear effect with a shallow slope, less than 45 degrees. If there is no effect, we should see a 45-degree line implying a uniform distribution of p-values (refer to our background information report, Young and Kindzierski 2018, Info 02).

We do see small p-values to the lower left in each image plot of Figure 1 and then points ascending in a roughly 45-degree line. We see bilinear plots – something that resembles hockey sticks, the blade of small p-value and the handle representing what look like random effects. In short, we see a heterogeneous set of p-values – a mixture of what could be real effects or the results of p-hacking on the lower left of each image plot, and what appear to be random effects on the right. This is evidence of heterogeneity. Each point within an image plot is for the same air component, e.g. ozone associated with heart attacks. These image plots indicate that many studies are without effects; these results conform to the Simonsohn et al. (2014) picture of researcher analysis flexibility, e.g. p-hacking. Further information about heterogeneous p-values is available to the reader in our background information report (Young and Kindzierski 2018, Info 05).

**Discussion**

There is now greater acknowledgement in published literature of likely causes of false published research (Pocock et al. 2004, Ioannidis 2008, Sarewitz 2012, Hubbard 2015, Chamber 2017, Harris 2017):

- Arbitrary data selection including sculpting datasets to fit a theory.
- Selection of a statistical model to support a point of view.
- Wide choice of methods for computing statistical analysis results.
- Arbitrary handling of covariates.
- Selective reporting of only "interesting" results, those that fit the story being advanced.

Journal peer reviewers – many of which are drawn from the same group of scientists that publish in a particular area – do not understand or are not willing ask probing questions about these practices. Many journals and their editors simply reject studies that do not show positive effects (publication bias) (Phillips 2004).

There are many incentives to seek p-values <0.05. There is a perceived need to have a p-value less than 0.05 to be published. A reported positive might elicit more funding and a negative finding is likely to stop funding to a researcher for the question at issue. If published scientific results are in one direction there is pressure to stay within the current science paradigm to get editorial and referee support. Researchers perceive that editors want a clean, simple story (refer to our background information report, Young and Kindzierski 2018, Info 06).

With large complex data sets and MTMM it can be technically easy to get p-values less than 0.05. Indeed, many of the p-values reported in our case study are less than 0.05. The fact that the p-value plots are bilinear hockey−stick like, shows that the MA is not measuring a homogeneous, overall effect. Ehm (2016) states as much:

> "In fact the very meaning of an overall effect size deserves consideration in the presence of substantial heterogeneity."

Regarding heterogeneity and the possibility of p-hacking in our case study, Mustafic et al. (2012) reported statistical heterogeneity ($I^2$) values for their overall analysis for the six air components – 83% (O3), 93% (CO), 71% (NO2), 65% (SO2), 57% (PM10) and 51% (PM2.5). Higgens et al. (2003) tentatively assigned low, moderate and high $I^2$ values of 25%, 50%, and 75% for MA indicating that analysis of all the air components in our case study had moderate to high heterogeneity.

There are many factors that could lead to heterogeneity. Some of the limitations of a MA where there is heterogeneity are worthwhile discussing. The main limitation is the type of heterogeneity. If heterogeneity is of 'degree only', then random weighting MA should be reasonable. By 'degree only', we mean that there is an underlying single etiology that can be modified up or down a bit. If heterogeneity is due to a mixture of etiologies, then any averaging scheme is averaging apples and oranges and is likely invalid. Our Figure 1 indicates a mixture of nominally significant studies with non-significant studies. These results are incompatible. The same hockey stick pattern (Figure 1) is observed elsewhere in multiple scientific areas (e.g., medical, nutritional, environmental epidemiology, environmental science, Young and Kindzierski, 2018, Info 07). Occam's Razor (simple explanation) supports p-hacking in our case study.

The Young et al. (2018) recent review of particulate matter−MI effects MA observed that base studies had the potential for massive multiple testing and multiple modelling with no statistical adjustments. Given what they and we (from our current case study) have learned, we offer further thoughts about potential limitations of a meta-analysis where there is heterogeneity and the possibility of p-hacking:

- Power could be low in studies with low sample size. However with our case study, inspection of many of the negative studies shown in Mustafic et al. (2012) Figures 1−3 had narrow confidence intervals/high power.

- Studies with non-significant results may not be reported (publication bias). It is known that positive studies are much more likely to be reported than negative studies at a ratio of about 10:1. It is possible that negative studies were not reported. Table 4 has a potential of 6x34=204 results, yet only 104 p-value results were reported. In some cases data may not have been available for some variables and it is possible that researchers did not report negative results for other variables in Table 4.
- Covariates could differ between the small p-value studies and the large p-value studies. The usual emphasis in a study is an over-all claim. Very loosely… *A causes B*. Covariates are used to reduce variability and take other real effects off the table, correcting for age or gender, for example. Suppose that the effect of a covariate is severe, what statisticians call *interaction*… *A causes B* if *C* is at one level but does nothing if at another. For example, *A causes B* for males, but not for females. It is possible that there is some unknown (even unmeasured) covariate that would produce a hockey stick pattern. However, as stated above we see the hockey stick pattern elsewhere in multiple scientific areas. There could be many unknown covariates, but a simple explanation for the most likely common cause of small p-values is p-hacking.
- Geography could different for the two groups in a bilinear p-value plot. Geographic heterogeneity for environmental epidemiology has been often reported (e.g., Cox 2017). If an air component is causal, it should be causal everywhere, but it is not. For example, Wang et al. (2015b) summarized studies from the search engine PubMed related to case-crossover investigations of particulate matter− MI effects published before 15 March 2015. Nineteen studies with greater than 1,000 MI events in eight different countries were identified; seven studies had negative effects. The possibility is that p-hacking is everywhere in positive studies (Head et al. 2015).

- Sample size/power is different across base studies. Here one would expect the larger p-values to be on a line extended from the small p-values in a p-plot. If the small p-values are to be taken as correct, e.g. <0.001, then Boos and Stefanski (2011) and Johnson (2013) argue for the reality of the effect. Here they are assuming that there is only one question at issue and good control in these studies. If the effect is judged real, then one should see some level of association in all well-conducted studies, i.e., all the p-values should be on a line with slope less than 45 degrees. On the other hand, if the p-value plot is bi-linear, it implies a mixture and p-hacking should be considered as an explanation for the small p-values.

- Sample designs are different across base studies. Randomized clinical trials (RCTs) work to a very different standard than do observational studies. If a MA has both RCTs and observational studies, seeing a hockey stick pattern in a p-value plot might not be surprising. However, Young and Karr (2011) evaluated 12 studies in which 52 claims coming from observational studies were tested in RCTs. They found that 100% of the observational claims (52 out of 52) failed to replicate. P-hacking should be considered as an explanation for lack of replication.

- Funding sources (researcher beliefs) could differ across base studies.

Regarding our findings and in understanding the reliability of base studies in MA, it is worth reconsidering the so-called file drawer problem. It is often difficult to publish a negative effect study, the argument being that it is easy to get a no-effect result. Editors today generally favour novel studies that show positive effects. It is known that researchers often put negative results in a "file drawer" and move on to other things (Hubbard 2015, Chamber 2017, Harris 2017). Conventional wisdom is that if one is doing MA, one should worry about unreported, negative results (Stroup et al. 2000, Ehm 2016). Because of the file drawer problem, there could be many "no effect" studies that are ignored and not being averaged with reported positive effect studies in MA.

Even if there are a lot of positive effect studies, does the file-drawer problem go away? A key contention of Simonsohn et al. (2014) is that researchers can get a small p-value by exploiting the flexibility of the data selection and analysis process. Parameters of a statistical analysis (analysis search space) can be varied until a p-value less than 0.05 is obtained, the number needed to be considered for publication. Simonsohn et al. (2014) proposed an analysis strategy to enable judgment about whether there is likely analysis manipulation in research when the only reported p-values are <0.05. They asserted that if observed significant p-values pile up at just below 0.05, then that is evidence of researcher analysis manipulation. This is very reasonable where an analysis search space is potentially large. We extend this reasoning to the situation where essentially all hypothesis testing p-value results are given. For example, with 34 papers and six outcomes used in our case study, there could be at least 34x6=204 tests of hypothesis. If confounders were considered, there could be a much larger number of statistical test results that should be reported.

So, are the reported small p-values in our case study the result of real effects or are they the result of p-hacking? Large, well-conducted studies (Milojevic et al. 2014, Young et al. 2017, Wang et al 2015b) find no association of air quality with heart health effects. Milojevic et al. (2014) reported evidence of no effects of air quality on a range of cardiovascular effects in England and Wales which supports that small p-values seen in our case study are the result of p-hacking, intentional or not. In our background information report (Young and Kindzierski 2018, Info 04) we extracted each estimated effect and its confidence limits from their study and computed p-values for 132 tests implied by their results (66 for cardiovascular disease (CVD) hospital admissions and 66 for CVD deaths). Volcano plots were constructed for MI hospital admissions and deaths (plots of the negative log10 of each p-value against effect size). A Bonferroni significance level reference line, -log10 (0.05/66) = 3.12 was used to identify the preponderance of significant effects for each end point. A PM10−non-MI admission effect was the only estimate significant after a Bonferroni correction (refer to Young and Kindzierski 2018, Info 04). As stated previously, our interpretation of resulting p-value and volcano plots constructed for the Milojevic et al. (2014) study is consistent with their finding that the air component− CVD effect is negative.

Young et al. (2017) recently examined the role of air quality and deaths in California. All death certificates for California were obtained for the years 2000-2012. There were over 2M death certificates, over 37,000 exposure days, and 8 air basins. After adjustment for seasonal effects, they observed no positive effect between PM2.5 or ozone (two primary air components of primary concern to public health) and acute mortality for 0- to 2-day lag effects.

Wang et al. (2015b) conducted two independent case-crossover studies of air quality (CO, NO, NO2, PM2.5 and ozone) and MI hospitalization events over the period 1999-2010 in two geographically-close and demographically similar sized cities of Calgary and Edmonton, Alberta, Canada (~1M population each) using a time-stratified design. Among 600 different statistical tests of potential air component−MI hospitalization effects investigated for the Calgary (Edmonton) population, only 1.17% (0.67%) had p-value less than 0.05. More importantly, none of the effects were reproduced in the two cities despite their geographic closeness (within 300 km of each other), and demographic and air quality similarities.

As we stated previously, many observational epidemiology studies support the current paradigm that certain air quality components are causal of effects such as death and MI from short-term exposure. However, in the presence of unmeasured confounders, such causality is difficult to establish using observational epidemiology for a relatively weak health risk factor such as ambient air quality (Cox 2017).

Within limits, observational epidemiology can be effective for looking at chronic diseases in populations and risk factors. However it is not easy to sort actual risk factors from statistical background noise of confounders and biases (which are inescapable elements of observational epidemiology). Current observational epidemiology methods allow one to tell a difference between effects from strong and weak risk factors. However, it is next to impossible using these methods to attempt to consistently differentiate between effects of weak risk factors and nothing at all, which is what is happening in ambient air quality−effects studies. Collectively, efforts chasing after relatively weak risk factors like ambient air quality mostly lead to outcomes that vibrate between showing effects/no effects dependent upon how researchers design studies, use data, analyze and report results (called 'vibration of effects' after Ioannidis 2008).

In looking closer at our case study, analysis of biomedical diagnostic test results being true depends on sensitivity, specificity and baseline prevalence of a disease in a population (Shaw 2003). Ioannidis (2005) extended the application of this by including bias for understanding the probability of a research claim being true for clinical trials, traditional epidemiological studies and modern molecular research. As the magnitude of the p-value is used by researchers for interpretation of statistical tests on observational data and when many statistical tests are performed simultaneously (a common feature of the base studies), the overall chance of a type I error (incorrect rejection of a true null hypothesis) can substantially exceed the nominal error rate used in each individual test (Parker and Rothenberg 1988, Westfall and Young 1993). Prevalence of individuals ≥18 years reporting a history of MI is statistically very low; for example in United States it is ~4% (CDC 2007). Because of this and based on the work of Ioannidis (2005), one would expect an unreasonably high number of false positive results in weak risk factor−MI observational studies with MTMM. Simple probability calculations support this. If n independent statistical tests are performed, each at the p=0.05 level, the probability of at least one erroneous significant finding is $1 - 0.95^n$ in the absence of bias (Selwyn 1989). For n=500 the overall probability is essentially 1; even at p=0.005 the overall probability is high (0.92).

As for strong MI risk factors, Yusef et al. (2004) conducted a standardized case-control study of population attributable acute MI risks in 52 countries in Asia, Europe, the Middle East, Africa, Australia and North and South America. This included the recruitment and study of 15,152 cases and 14,820 controls. Structured questionnaires were administered, physical examinations were undertaken and non-fasting blood samples were drawn from each participant. Based on their data, Yusef et al. (2004) estimated that nine predisposing risk factors account for the majority (90%) of population attributable acute MI risks in men (94% in women) – abnormal lipids, smoking, hypertension, diabetes, abdominal obesity, psychosocial factors, consumption of fruits and vegetables, alcohol consumption and regular physical activity. Air quality did not rise to the level of importance to be included as a risk factor in their study.

Air quality−MI observational studies are population-based. Covariates tend to include age, sex, co-morbidities, co-pollutants and weather variables (e.g., temperature, relative humidity) (Wang et al. 2015a,b). This was a common feature of the base studies; however treatment of covariates differed among the studies. This partially explains the wide variation in number of statistical tests possible across the base studies (Table 3)… *the many possible different averaging times that can be used to represent an air component or a weather variable and the greater the number of covariates considered, the greater the number of tests possible*. This is a disadvantage of combining observational studies and it adds a layer of uncertainty to interpretation of MA statistical results.

Putting this all together, a practical mistake that researchers can make when using statistics – particularly for studying weak risk factors – is not recognizing that the data being analyzed is insufficient to answer a research question if one does not understand its limitations. By this we mean limitations in the data and methods being used. When one finds a statistically significant result using MA, one is truly in a statistical world not the real world. Low MI prevalence, numerous strong MI risk factors unaccounted for, inconsistent averaging of predictors, inconsistent handling of covariates and possible p-hacking in the base studies offers evidence for necessary caution in the interpretation of the MA case study results.

**Summary and conclusions**

Findings of our case study, a highly cited study in scientific literature – that the air components CO, NO2, SO2, PM10, PM2.5 except for ozone are significantly associated with a near-term increase in health attack risk – are not supported by our analysis of the base studies. P-values used in the case study are heterogeneous and the small p-values have the appearance of being researcher-generated rather than unbiased evidence of real effects. We conclude that this meta-analysis does not provide reliable evidence for an association of air quality components with myocardial risk. Because of this, we recommend that a need exists to further examine the reliability of other meta-analysis applications of air quality−short-term health effects using independent statistical methods demonstrated here or other suitable statistical methods.

Understanding the burden of chronic disease and death is of importance to public health policy makers due to aging populations in North America, Western Europe and elsewhere. Making sensible changes in public health policy is key for promoting and protecting health of our aging populations. The best available scientific evidence is needed to guide public health policy makers in this effort. Our findings suggest the appearance of heterogeneous, researcher-generated p-values used in meta-analysis of air quality–heart attack risk. These types of analysis can lead to false published evidence with the potential for further misuse of this evidence by public health policy makers. The extent of this problem is unknown and warrants further investigation. Most importantly, policy makers – where necessary – need to be cautioned about using air quality−effect meta-analysis results where some of the base papers show evidence consistent with p-hacking.

**Acknowledgements**

The authors gratefully acknowledge comments provided by three reviewers selected by the Editor and anonymous to the authors. The comments helped improve the quality of the final article. The authors would also like to thank the private not-for-profit Cornell University Library e-print service repository arXiv.org for allowing background information related to this paper to be posted online. No other individuals or organizations contributed to the quality and content of the paper.

**Declaration of interest**

Dr. Young was self-employed with CGStat, Raleigh, North Carolina during the writing of this manuscript. CGStat is a for-profit company owned by Dr. Young and it provides independent statistical consulting to clients in United States. He previously worked at Eli Lilly, GlaxoSmithKline, and the National Institute of Statistical Sciences. He is a member of the Scientific Advisory Board of the U.S. Environmental Protection Agency. In the past he has received grants from the American Petroleum Institute for work in this area; however no financial support was provided for the writing of this paper. Dr. Kindzierski was employed in the School of Public Health at the University of Alberta, Edmonton, Alberta during the writing of this paper (University of Alberta Continuing Academic Staff, calendar.ualberta.ca/content.php?catoid=28&navoid=7162). Dr. Kindzierski retired in fall 2018. He has received funding from Public Health Agency of Canada, Saskatchewan Health, Alberta Environment and Parks, Canada's Oil Sands Innovation Alliance (COSIA), EPCOR Water Services, TransAlta Corporation and the not-for-profit West Central Airshed Society in the last five years. During this time he has not participated in any capacity in litigation or regulatory proceedings or an advocacy role on behalf of any organization related to the contents of the paper. No financial support was provided to Dr. Kindzierski for the writing of this paper. The research undertaken and preparation of the paper was the professional work product of the authors. Both authors have sole responsibility for the writing and content of the paper. The work has not been influenced by anyone other than the authors and the paper was not shared with any other individuals during its preparation. The analysis performed and the conclusions drawn are exclusively those of the authors. The authors report that no financial interest or benefit will arise from the direct application of this research.

**Supplementary material**

Supplemental information for this article is provided at the end. Background information on meta-analysis evaluation is freely available at https://arxiv.org/abs/1808.04408.
25

**ORCID**

S. Stanley Young	https://orcid.org/0000-0001-9449-5478

Warren B. Kindzierski	http://orcid.org/0000-0002-3711-009X

modifiable risk factors associated with myocardial infarction in 52 countries (the INTERHEART study): case-control study. Lancet. 364:937−952. doi:10.1016/S0140-6736(04)17018-9.

Table 1. Risk ratios and Confidence Limits for six air components.

| Air Component | Risk Ratio | Cl low[a] | Cl high[b] |
|---|---|---|---|
| carbon monoxide | 1.048 | 1.026 | 1.070 |
| nitrogen dioxide | 1.011 | 1.006 | 1.016 |
| sulfur dioxide | 1.010 | 1.003 | 1.017 |
| PM10 | 1.006 | 1.002 | 1.009 |
| PM2.5 | 1.025 | 1.015 | 1.036 |
| ozone | 1.003 | 0.997 | 1.010 |

[a]Cl Low = lower confidence limit; [b]Cl High = upper confidence limit.

Table 2. Authors, variable counts, and analysis search spaces for the 34 case study base papers.

| Cit # | Author | Outcome | Predictor | Covariate | Lag | Space1 | Space2 | Space3 |
|---|---|---|---|---|---|---|---|---|
| 7 | Braga | 4 | 1 | 6 | 4 | 16 | 64 | 1,024 |
| 8 | Koken | 5 | 5 | 6 | 5 | 125 | 64 | 8,000 |
| 9 | Barnett | 7 | 5 | 10 | 1 | 35 | 1,024 | 35,840 |
| 10 | Berglind | 1 | 4 | 10 | 2 | 8 | 1,024 | 8,192 |
| 11 | Cendon | 2 | 5 | 5 | 8 | 80 | 32 | 2,560 |
| 12 | Linn | 3 | 4 | 8 | 3 | 36 | 256 | 9,216 |
| 19 | Ye | 8 | 5 | 3 | 5 | 200 | 8 | 1,600 |
| 20 | Peters | 1 | 8 | 11 | 2 | 16 | 2,048 | 32,768 |
| 21 | Rich | 1 | 5 | 9 | 6 | 30 | 512 | 15,360 |
| 22 | Sullivan | 4 | 4 | 8 | 3 | 48 | 256 | 12,288 |
| 23 | Eilstein | 1 | 12 | 5 | 6 | 72 | 32 | 2,304 |
| 24 | Lanki | 1 | 5 | 3 | 6 | 30 | 8 | 240 |
| 25 | Maté | 4 | 6 | 7 | 6 | 144 | 128 | 18,432 |
| 26 | Medina | 15 | 6 | 8 | 6 | 540 | 256 | 138,240 |
| 27 | Poloniecki | 7 | 5 | 5 | 1 | 35 | 32 | 1,120 |
| 28 | Stieb | 6 | 6 | 7 | 3 | 108 | 128 | 13,824 |
| 29 | Zanobetti | 5 | 2 | 11 | 3 | 30 | 2,048 | 61,440 |
| 30 | Zanobetti | 5 | 18 | 8 | 3 | 270 | 256 | 69,120 |
| 31 | Zanobetti | 5 | 2 | 9 | 2 | 20 | 512 | 10,240 |
| 32 | Hoek | 4 | 8 | 9 | 3 | 96 | 512 | 49,152 |
| 33 | Cheng | 1 | 5 | 6 | 3 | 15 | 64 | 960 |
| 34 | Hsieh | 1 | 5 | 6 | 3 | 15 | 64 | 960 |
| 35 | Pope | 1 | 2 | 13 | 7 | 14 | 8,192 | 114,688 |
| 36 | D'Ippoliti | 3 | 4 | 11 | 3 | 36 | 2048 | 73,728 |
| 37 | Henrotin | 4 | 5 | 14 | 14 | 280 | 16,384 | 4,587,520 |
| 38 | Ueda | 3 | 1 | 7 | 3 | 9 | 128 | 1,152 |
| 39 | Mann | 4 | 4 | 9 | 7 | 112 | 512 | 57,344 |
| 40 | Sharovsky | 4 | 3 | 10 | 8 | 96 | 1,024 | 98,304 |
| 41 | Belleudi | 4 | 3 | 8 | 13 | 156 | 256 | 39,936 |
| 42 | Nuvolone | 1 | 3 | 9 | 8 | 24 | 512 | 12,288 |
| 43 | Peters | 4 | 5 | 10 | 4 | 80 | 1,024 | 81,920 |
| 44 | Ruidavets | 4 | 3 | 8 | 4 | 48 | 256 | 12,288 |
| 45 | Zanobetti | 2 | 6 | 7 | 3 | 36 | 128 | 4,608 |
| 46 | Bhaskaran | 1 | 5 | 7 | 5 | 25 | 128 | 3,200 |

Note: Cit # before author name is the case study reference number; author name is first author listed (refer to our Supplement).

Table 3. Summary statistics for the number of possible analyses using the three search spaces.

| Statistic | Space1 | Space2 | Space3 |
|---|---|---|---|
| maximum | 540 | 16,384 | 4,587,520 |
| quartile | 109 | 1,024 | 58,368 |
| median | 36 | 256 | 12,288 |
| quartile | 23 | 64 | 2,496 |
| minimum | 8 | 8 | 240 |

Table 4. P-values for each paper and air component reported in the case study and plotted in Figure 1.

| Cit# | Author | ozone | CO | NO2 | SO2 | PM10 | PM2.5 |
|---|---|---|---|---|---|---|---|
| 7 | Braga | | | | | 0.001 | |
| 8 | Koken | *0.001* | | | | | |
| 9 | Barnett | | 0.19 | 0.25 | | | 0.21 |
| 10 | Berglind | *0.42* | *0.65* | *0.65* | | *0.9* | |
| 11 | Cendon | 0.02 | *0.95* | 0.32 | 0.001 | 0.24 | |
| 12 | Linn | *0.31* | 0.001 | 0.03 | | 0.001 | |
| 19 | Ye | | | 0.001 | | | |
| 20 | Peters | *0.78* | *0.91* | 0.39 | *0.54* | 0.01 | 0.009 |
| 21 | Rich | | | | | | 0.66 |
| 22 | Sullivan | | 0.55 | | 0.99 | | 0.38 |
| 23 | Eilstein | 0.04 | 0.13 | 0.31 | 0.21 | | |
| 24 | Lanki | *0.14* | 0.05 | *0.35* | | 0.46 | |
| 25 | Maté | | | | | | 0.001 |
| 26 | Medina | 0.02 | | 0.08 | 0.02 | *0.45* | |
| 27 | Poloniecki | *0.19* | 0.005 | 0.009 | 0.001 | | |
| 28 | Stieb | *0.12* | 0.05 | *0.79* | 0.09 | 0.77 | 0.25 |
| 29 | Zanobetti | | | | | | |
| 30 | Zanobetti | | | | | | 0.001 |
| 31 | Zanobetti | | | | | | 0.002 |
| 32 | Hoek | 0.33 | 0.03 | 0.1 | 0.33 | 0.7 | |
| 33 | Cheng | 0.001 | 0.001 | 0.008 | 0.33 | 0.13 | |
| 34 | Hsieh | 0.001 | 0.001 | 0.001 | 0.14 | 0.001 | |
| 35 | Pope | | | | | | 0.04 |
| 36 | D'Ippoliti | | 0.21 | 0.62 | | | |
| 37 | Henrotin | *0.3* | | | | | |
| 38 | Ueda | | | | | | 0.24 |
| 39 | Mann | *0.009* | 0.001 | 0.001 | | 0.87 | |
| 40 | Sharovsky | | 0.16 | | 0.02 | 0.84 | |
| 41 | Belleudi | | | | | 0.09 | 0.04 |
| 42 | Nuvolone | | 0.16 | 0.04 | | 0.55 | |
| 43 | Peters | *0.01* | 0.08 | *0.76* | 0.02 | | 0.07 |
| 44 | Ruidavets | 0.003 | | *0.76* | 0.91 | | |
| 45 | Zanobetti | *0.36* | 0.04 | 0.001 | | 0.001 | 0.02 |
| 46 | Bhaskaran | 0.74 | *0.48* | *0.22* | *0.6* | *0.12* | |

Note: Cit# is the case study citation number; author is first author listed; blank cell indicates no results or results were not used for MA by Mustafic et al. (2012); p-values in *Italics* are associated with relative risks that less than 1.

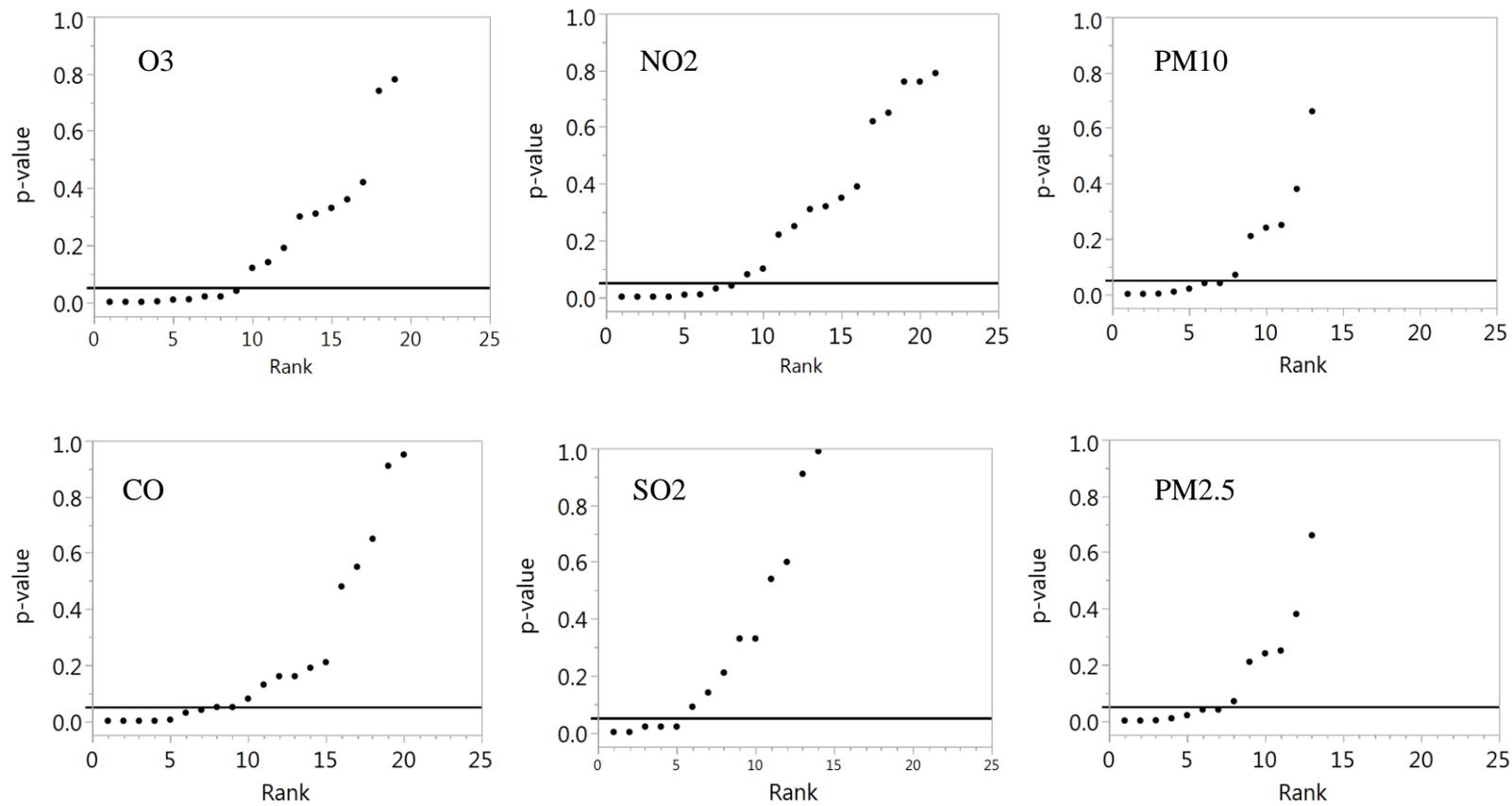

Figure 1. P-value plots for six air quality components of the case study (horizontal black line is *p*=0.05).

# Supplement

# Evaluation of a meta-analysis of air quality and heart attacks, a case study

S. Stanley Young[a] and Warren B. Kindzierski[b,c]

[a]*CGStat, Raleigh, NC, USA;* [b]*School of Public Health, University of Alberta, Edmonton, Alberta, Canada*

[c]Corresponding author: School of Public Health, University of Alberta, 3-57 South Academic Building, 11405-87 Avenue, Edmonton, Alberta, T6G 1C9 Canada; warrenk@ualberta.ca

**11 Pages, 1 Table**

## Case Study Selection Strategy

For the selection of our case study, we performed a search using the Web of Science electronic database (Clarivate Analytics, Philadelphia, PA) within the University of Alberta libraries system (www.library.ualberta.ca) on 28 June 2018.

Web of Science (formerly ISI Web of Knowledge) is an online subscription-based scientific citation indexing service of multiple databases that reference cross-disciplinary research. Web of Science includes over 50,000 scholarly books, 12,000 journals and 160,000 conference proceedings.

We searched the Web of Science records between the period 1 January 1980 and 28 June 2018 using the following strategy:

- An initial search was performed using the terms *meta-analysis* AND *myocardial infarction* within a record title. This search yielded 1,024 results.
- A second independent search was performed using the terms *air pollutants* OR *air pollution* within a record title. This search yielded 36,239 results.
- A combined search of initial and secondary results was then performed. This search yielded 3 results.

A screenshot image of the Web of Science search history results is shown below:

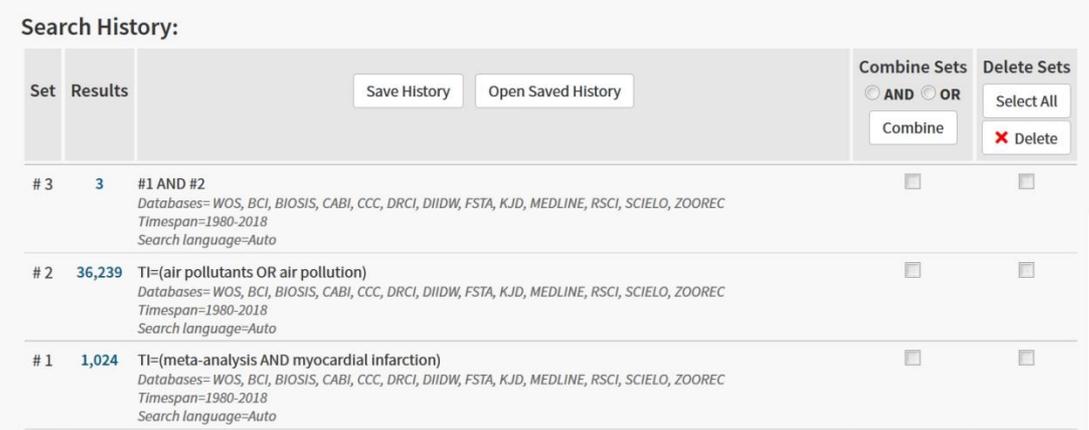

The Web of Science record for set #3 results was:

1. Main Air Pollutants and Myocardial Infarction A Systematic Review and Meta-analysis

By: Mustafic, Hazrije; Jabre, Patricia; Caussin, Christophe; et al.

JAMA-JOURNAL OF THE AMERICAN MEDICAL ASSOCIATION

Volume: 307   Issue: 7   Pages: 713-721

Published: FEB 15 2012

Times Cited: 226 (from All Web of Science Databases)[a]

[a]  Web of Science note – *As of March/April 2018, this highly cited paper received enough citations to place it in the top 1% of its academic field based on a highly cited threshold for the field and publication year.*

2. Short-term exposure to particulate air pollution and risk of myocardial infarction: a systematic review and meta-analysis

By: Luo, Chunmiao; Zhu, Xiaoxia; Yao, Cijiang; et al.

ENVIRONMENTAL SCIENCE AND POLLUTION RESEARCH

Volume: 22   Issue: 19   Pages: 14651-14662

Published: OCT 2015

Times Cited: 11 (from All Web of Science Databases)

3. Air Pollution and Myocardial Infarction: A Systematic Review and Meta-Analysis

By: Mustafic, Hazrije; Jabre, Patricia; Caussin, Christophe; et al.

CIRCULATION

Volume: 124   Issue: 21   Supplement: S   Meeting Abstract: A11876

Published: NOV 22 2011

Times Cited: 0 (from All Web of Science Databases)

The 1st study (Main Air Pollutants and Myocardial Infarction A Systematic Review and Meta-analysis) had the highest citation record and was selected as the case study.

**References of 34 Base Papers used in Case Study (number indicated on the left is the reference number in Mustafic et al. (2012))**

**Table S1  Summary description of Mustafic et al. (2012) base studies.**

| Cit #[1] | Location | Time period | Data source | MI events | Air pollutants | Study type | Model type | Study quality[2] |
|---|---|---|---|---|---|---|---|---|
| 7 | 10 US cities | 1986−1993 | Death registry | Not given | PM10 | Time-series | Mono-pollutant | Good |
| 8 | Denver, US | 1993−1997 | MI hospital admissions | 1,576 | O3, CO, NO2, SO2, PM10 | Time-series | Mono-pollutant | Low |
| 9 | Australia (5 cities), New Zealand (2 cities) | 1998−2001 | MI hospital admissions | Not given | O3, CO, NO2, PM10, PM2.5 | Case crossover | Mono-pollutant & multi-pollutant | Good |
| 10 | Stockholm, Sweden | 2001−2007 | MI registry | 660 | O3, CO, NO2, PM10 | Case crossover | Mono-pollutant | Intermediate |
| 11 | Sao Paulo, Brazil | 1998−1999 | MI hospital admissions | 19,272 | O3, CO, NO2, SO2, PM10 | Time-series | Mono-pollutant | Low |
| 12 | Los Angeles, US | 1988−1994 | MI hospital admissions | Not given | O3, CO, NO2, PM10 | Time-series | Mono-pollutant | Intermediate |
| 19 | Tokyo, Japan | 1980−1995 | MI emergency hospital admissions | Not given | O3, CO, NO2, SO2, PM10 | Time-series | Mono-pollutant | Low |
| 20 | Boston, US | 1999−2001 | MI registry | 772 | O3, CO, NO2, SO2, PM10, PM2.5 | Case crossover | Mono-pollutant | Low |
| 21 | New Jersey, US | 2004−2006 | MI hospital admissions | 5,864 | PM2.5 | Case crossover | Multi-pollutant | Intermediate |
| 22 | Washington, DC | 1988−1994 | MI hospital admissions | 5,793 | CO, NO2, PM2.5 | Case crossover | Mono-pollutant | Intermediate |
| 23 | Strasbourg, France | 1984−1989 | MI registry | Not given | O3, CO, NO2, SO2 | Time-series | Mono-pollutant | Good |

[1] Citation number of Mustafic et al. (2012) base study.
[2] General quality rating of study assigned by Mustafic et al. (2012).

**Table S1  Summary description of Mustafic et al. (2012) base studies (con't).**

| Cit #[1] | Location | Time period | Data source | MI events | Air pollutants | Study type | Model type | Study quality[2] |
|---|---|---|---|---|---|---|---|---|
| 24 | Europe (5 cities) | 1992−2000 | MI registry & MI hospital admissions | 26,854 | O3, CO, NO2, PM10 | Time-series | Mono-pollutant | Good |
| 25 | Madrid, Spain | 2003−2005 | Death registry | 1,096 | PM2.5 | Time-series | Mono-pollutant | Good |
| 26 | Paris, France | 1991−1995 | Registry of doctor's house calls | Not given | O3, NO2, SO2, PM10 | Time-series | Mono-pollutant | Low |
| 27 | London, UK | 1987−1994 | MI hospital admissions | 68,300 | O3, CO, NO2, SO2 | Time-series | Mono-pollutant & multi-pollutant | Good |
| 28 | 14 cities in Canada | 1992−2003 | MI emergency hospital admissions | 63,184 | O3, CO, NO2, SO2, PM10, PM2.5 | Time-series | Mono-pollutant | Low |
| 29 | 21 US cities | 1986−1999 | MI hospital admissions | 302,245 | PM10 | Case crossover | Mono-pollutant | Intermediate |
| 30 | 112 US cities | 1999−2005 | Death registry | 397,894 | PM2.5 | Time-series | Mono-pollutant | Intermediate |
| 31 | 26 US cities | 2000−2003 | MEDICARE registry & MI hospital admissions | 121,652 | PM2.5 | Time-series | Mono-pollutant | Intermediate |
| 32 | Netherlands | 1986−1994 | Death registry | 62 per day | O3, CO, NO2, SO2, PM10 | Time-series | Mono-pollutant | Good |
| 33 | Kaohsiung, Taiwan | 1996−2006 | MI hospital admissions | Not given | O3, NO2, SO2, PM10 | Case crossover | Mono-pollutant & multi-pollutant | Low |
| 34 | Taipei, Taiwan | 1996−2006 | MI hospital admissions | 23,420 | O3, CO, NO2, SO2, PM10 | Case crossover | Mono-pollutant | Low |

[1] Citation number of Mustafic et al. (2012) base study.
[2] General quality rating of study assigned by Mustafic et al. (2012).

**Table S1  Summary description of Mustafic et al. (2012) base studies (con't).**

| Cit #[1] | Location | Time period | Data source | MI events | Air pollutants | Study type | Model type | Study quality[2] |
|---|---|---|---|---|---|---|---|---|
| 35 | Utah, US | 1991−2001 | Angiographic registry | 3,910 | PM2.5 | Case crossover | Mono-pollutant | Intermediate |
| 36 | Roma, Italy | 1995−1997 | MI hospital admissions | 6,531 | CO, NO2 | Case crossover | Mono-pollutant | Good |
| 37 | Dijon, France | 2001−2007 | MI registry | 771 | O3 | Case crossover | Multi-pollutant | Good |
| 38 | 9 cities in Japan | 2002−2004 | Death registry | 67,897 | PM2.5 | Time-series | Mono-pollutant | Low |
| 39 | California, US | 1988−1995 | Insurance registry | 19,690 | O3, CO, NO2, PM10 | Time-series | Mono-pollutant | Intermediate |
| 40 | Sao Paulo, Brazil | 1996−1998 | Death registry | 12,007 | CO, SO2, PM10 | Time-series | Mono-pollutant | Good |
| 41 | Roma, Italy | 2001−2005 | MI emergency hospital admissions | 22,659 | PM10, PM2.5 | Case crossover | Mono-pollutant | Intermediate |
| 42 | Tuscany, Italy | 2002−2005 | MI registry | 11,450 | CO, NO2, PM10 | Case crossover | Mono-pollutant & multi-pollutant | Intermediate |
| 43 | Augsburg, Germany | 1999−2001 | MI registry | 851 | O3, CO, NO2, SO2, PM10, PM2.5 | Case crossover | Mono-pollutant | Low |
| 44 | Toulouse, France | 1997−1999 | MI registry | 399 | O3, NO2, SO2 | Case crossover | Mono-pollutant | Good |
| 45 | Boston, US | 1995−1999 | MI hospital admissions | 15,578 | O3, CO, NO2, PM2.5 | Case crossover | Mono-pollutant | Good |
| 46 | England & Wales | 2003−2006 | MI registry | 79,288 | O3, CO, NO2, PM2.5 | Case crossover | Mono-pollutant & multi-pollutant | Good |

[1] Citation number of Mustafic et al. (2012) base study.
[2] General quality rating of study assigned by Mustafic et al. (2012).